# Analysis of a Failed Eclipse Plasma Ejection Using EUV Observations

**E. Tavabi[1,2] . S. Koutchmy[1] . C. Bazin[1,3]**

**Abstract** The photometry of eclipse white-light (W-L) images showing a moving blob is interpreted for the first time together with observations from space with the *PRoject for On Board Autonomy* (PROBA-2) mission (ESA). An off-limb event seen with great details in W-L was analyzed with the SWAP imager (*Sun Watcher using Active pixel system detector and image Processing*) working in the EUV near 174 Å. It is an elongated plasma blob structure of 25 Mm diameter moving above the east limb with coronal loops under. Summed and co-aligned SWAP images are evaluated using a 20 hour sequence, in addition to the 11 July, 2010 eclipse W-L images taken from several sites. The *Atmospheric Imaging Assembly* (AIA) instrument on board the *Solar Dynamics Observatory* (SDO) recorded the event suggesting a magnetic reconnection near a high neutral point; accordingly, we also call it a magnetic plasmoid. The measured proper motion of the blob shows a velocity up to 12 km $s^{-1}$. Electron densities of the isolated condensation (cloud or blob or plasmoid) are photometrically evaluated. The typical value is $10^8$ $cm^{-3}$ at $r$=1.7 $R_\odot$, superposed on a background corona of $10^7$ $cm^{-3}$ density. The mass of the cloud near its maximum brightness is found to be $1.6\times10^{13}$ gr which is typically $0.6\times10^{-4}$ of the overall mass of the corona. From the extrapolated magnetic field the cloud evolves inside a rather broad open region but decelerates, after reaching its maximum brightness. The influence of such small events for supplying material to the ubiquitous slow wind is noticed. A precise evaluation of the EUV photometric data after accurately removing the stray light, suggests an interpretation of the weak 174 Å radiation of the cloud as due to resonance scattering in the Fe IX/X lines.

E. Tavabi
tavabi@iap.fr

S. Koutchmy
koutchmy@iap.fr

C. Bazin
bazin@iap.fr

[1] Institut d'Astrophysique de Paris, Sorbonne Universités- UMR 7095, CNRS and UPMC, F-75014 (France), koutchmy@iap.fr,
[2] Physics Department, Payame Noor University (PNU), 19395-3697, Tehran, I. R. of Iran,
[3] UPMC, LISE laboratory, 4 Place Jussieu 75252 Paris.

## 1. Introduction

The modeling of the solar wind (SW) is classically given in the framework of a kinetic and a fluid approach (see *e.g.* Echim *et al.,* 2011). Several decades ago when coronal holes were identified the origin of the fast wind was established. Shortly after coronal transients called today coronal mass ejections (CMEs) started to be studied. The remaining slow stationary solar wind is now the subject of intensive study, see *e.g.* Wang (2011), Antiochos *et al.* (2011), Goryiev *et al.* (2014). W-L observations taken with the *Large Angle and Spectrometric Coronagraph* (LASCO) instrument on board the *Solar and Heliospheric Observatory* (SoHO) C2 coronagraph, and further out with the C3 coronagraph, were assembled to make accelerated movies at the time of the solar minimum (1997) of Cycle 22. They convincingly showed persistent and rather impressive flows along streamers. Difference movies made later show the moving features called blobs flying across the FOV (Sheeley *et al.,* 1997, Wang *et al.,* 2000) qualitatively confirming that streamers are a potential source for the slow wind (see also Koutchmy and Livshits 1992) although the suggestion of streamers being a part of the heliosheet with a simplified anti-parallel radial magnetic field makes the process not evident unless small plasmoids are produced. Radially moving plasma blobs seen above $r$=2.5 R$_\odot$ with velocities over 100 kms$^{-1}$ were responsible for a dynamic picture as blobs are considered to be tracers of the SW. It was noted that blobs could not be found in the LASCO (SoHO) C1 coronagraph FOV using the 530.3 nm Fe XIV filtergrams (Sheeley *et al.,* 1997). In his recent review Y-M. Wang (2011) considers the mechanisms producing both fast and slow winds as closely related, under the general assumption that they are the result of magnetic reconnections occurring at various heights above the surface. In the case of the slow wind, he indicates that both the so-called cusps of streamers and edge of streamers produce a portion of the slow wind. The rest of the slow wind comes from interactions with the neighboring coronal holes located near active regions and from just inside the polar hole boundaries, where the outflow is driven by energy deposition at low heights along open lines.

We first note that these sources of the slow wind are well below the typical radial distance (of order of $r$=2.5 R$_\odot$) where the occulting disk of the LASCO C2 coronagraph completely masks the most massive part of the corona. Therefore the sources of the SW are not imaged. EUV imagers, on the other hand can be used for the analysis of the corona on and very near the solar disk; they however reflect the variations of both the temperature and the plasma density.

The internal occulting disk of the more recent COR1 coronagraphs of the STEREO mission is much smaller than 2.5 R$_\odot$; their effective spatial resolution and signal/noise ratio are just sufficient to record the small plasma blobs detected in its FOV using the linear polarization analysis. They were called localized plasma density enhancements by Jones and Davila (2009) who deduced in W-L similar properties to the LASCO C2 blobs, but starting from 1.5 R$_\odot$ and with radial speeds above 50 kms$^{-1}$. No attempt to identify the events with a EUV counterpart was yet reported.

Aside from the flares that produce surges and other transient phenomena, it is difficult to claim that the sources of the slow wind have been identified with EUV or even SXR imagers, although the SPIRIT telescope aboard *Complex ORbital Observations Near-Earth of Activity of the Sun* (CORONAS-F) with its large FOV (Slemzin *et al.,* 2008; 2011), demonstrated that above some mid-latitude active regions EUV coronal rays, possibly indicating plasma outflows, are recorded out to distances of 2 to 3 R$_\odot$. A long LASCO C2 W-L ray was studied in Filippov *et al.* (2011) with a specific quadrupolar configuration evidenced in EUV emissions and closer to what is proposed for SXR jets. Tsuneta (1997) described a rising plasmoid moving at 98 kms$^{-1}$ from a compact hot region where a M3.6 flare occurred. HXR were recorded and the temperature of the plasmoid was estimated to be of order of 15 MK, but no discussion was given of the destiny of this plasmoid as the event was mainly observed at low levels and, unfortunately, nothing is known above $r$=1.33 R$_\odot$.

Diamagnetic plasmoids were theoretically proposed to explain the solar wind, *e.g.* by Pneuman (1983) and later Mullan (1990). Sources of plasmoids were considered to be the result of reconnection occurring near the surface, producing coronal detached “spikes” or giant spicules moving away (Koutchmy and Loucif, 1991). Today the mechanism has been applied to the case of CH (*e.g.* Filippov *et al.,* 2009), *i.e.* for explaining jets and possibly the fast wind. Coronal streamers were excluded although some SXT studies report the observation near the boundaries of some active regions of sporadic high-speed outflow which may play an important role in the mass loading process of the slow solar wind (Guo *et al.,* 2010).

The myriads of coronal loops populating the low corona demonstrate that a lot of plasma is stored in the magnetic fields rooted in the photospheric layers; but higher up, where the plasma could eventually escape, very few observations exist. Only hot jets were observed in the low corona, first in SXR with Yohkoh (Yokoyama and Shibata, 1996), then with the SXT of Hinode (*e.g.* Filippov *et al.,* 2009). Not surprisingly, they are now observed in EUV lines and even in cool chromospheric lines, permitting a bridge with the former observations of surges, sprays and even macro-spicules and spicules (Koutchmy and Loucif, 1991). These confined eruptive fast phenomena with a typical Eiffel tower shape do not reach the critical radial distance from which they could escape. However, an interesting by-product of the 2D simulation of the magnetic reconnection performed by Yokoyama and Shibata (1996), was to predict, in addition to hot jets, the appearance of cool plasmoids injected into the surrounding corona as a result of the fast reconnection processes occurring near the top of the emerging loops. Both the plasmoids and the fast hot jet move along the field lines of the magnetic field. Following the early suggestion of Syrovatskii (1982) and the seminal simulation work of Mikic and Linker (1994), more sophisticated 2.5D MHD modeling were performed recently (*e.g.* Linker *et al.,* 2011; Antochios *et al.,* 2011; Zuccarello *et al.,* 2011; Bemporad *et al.,* 2011; Amari *et al.,* 2015), eventually based on strong shearing of the underlying magnetic field; they sometimes predict a plasmoid-like phenomenon (Bemporad *et al.,* 2011) although their motivation is basically to produce a CME event in achieving a full dynamic simulation.

CMEs are highly non-stationary phenomena associated with flares and/or filament eruptions that cannot explain the quasi-permanent flow of the slow wind. Mini-CMEs studied by Podladchikova *et al.* (2010) using the STEREO mission EUV filtergrams could perhaps supply mass to the slow wind. Very recently, again based on the use of the STEREO mission filtergrams, studies of a new type of CME, the so-called stealth CMEs appeared in the literature (see Ma *et al.,* 2010 for a survey at the time of solar minimum), but no extended statistical analyses for different periods of the solar cycle are yet available to make a comparison with the properties of the slow wind. They are probably good candidates to be a sporadic component of the wind, (*e.g.* Robbrecht *et al.,* 2009-a) and they will certainly be the subject of more studies in the future. The important point here is that it has been possible, thanks to the new STEREO EUV imagers, to look further out in the corona, compared to what was done before. The possibility of observing the EUV emissions of the far off-limb corona, out to 3 solar radius, was first demonstrated using the EUV Spirit imager of the Coronas/F mission at solar maximum (Slemzin *et al.,* 2008).

Finally, solar total eclipses provide wide field snapshots where detached features were episodically identified in W-L when the spatial resolution was sufficient, (see *e.g.* Koutchmy *et al.,* 1973; 1996; Zhukov *et al.,* 2000 for a photometric analysis of eclipse plasmoids). In Vial *et al.* (1992) ; Delannée *et al.* (1998) and Zhukov *et al.* (2000) a small, cool and short live eclipse plasmoid was analyzed with great details at a distance of 90 Mm from the surface, showing proper motion of 40 to 80 kms$^{-1}$, before fading into the surrounding corona. However, it was a really very small plasma cloud of 1.5 Mm size which is of little significance in the corona for the mass loading of the whole corona point of view. In W-L the K corona is revealed by Thomson scattering and a detailed photometric analysis permits one to deduce the electron densities, provided an absolute calibration is available for the coronal brightness to be expressed in units of the average solar disk brightness (Koutchmy *et al.,* 2012; November and Koutchmy, 1996). In the case of a small feature, some geometric assumptions are needed in order to disentangle the line of sight integration effect. The polarization analysis of the isolated in space coronal feature can help (Koutchmy, 1972) or alternatively, a pseudo-stereo view is used with observations coming from the SECCHI imagers of the STEREO mission. The addition of simultaneously obtained EUV filtergrams also provides an invaluable tool for interpreting W-L observations, especially in the case of a slowly moving cloud-like feature of long lifetime.

Based on the absolute photometric analysis of streamers, from the near surface heights up to 6 to 10 solar radius performed on eclipse images, it has been possible to measure the radial variation of the electron densities inside the streamers and even the distribution across (Koutchmy, 1972). From simplifying assumptions i) a Gaussian density distribution across the streamer with an elliptical cross-section suggesting a sheet- like structure; ii) plasma made of protons, electrons and alpha particles only; iii) hydrostatic temperature measured under 1 solar radius height and the iso-thermal assumption higher; iv/ a quasi- stationarity of dynamical processes and using the equation of mass continuity applied to the whole large scale structure, it has readily been possible to deduce the radial variation of the flow velocity inside the streamer (see Koutchmy, 1972; Koutchmy and Livshits, 1992). Typical values roughly agree with the predictions of Parker's model for the temperature measured in the low parts of the streamer, under 2 solar radius, and interestingly enough, very low flow velocities were also deduced in these parts under 2 $R_\odot$, with values less than 10 kms$^{-1}$ (Koutchmy, 1972). These values are smaller than the typical so- called turbulent velocity shown by line profiles integrated in the corona, which is of order of 20 kms$^{-1}$. We finally note that modern numerical 3D MHD modeling of the stretching by the solar wind of magnetic coronal structures now includes the magnetic field in a self- consistent way (Mikic *et al.,* 1999) without showing the formation of plasmoids ("blobs"), but with flows included.

## 2. Observations and Results

At the solar total eclipse of 11 July, 2010 the corona was recorded from sites located on the islands of the Pacific ocean (see Pasachoff *et al.,* 2011; Habbal *et al.,* 2011; Koutchmy *et al.,* 2012). Figure 1 shows a composite in polar coordinates made from the reconstructed eclipse W-L image taken by our main team on the Hao atoll (French Polynesia), Koutchmy *et al.,* (2012) with the SWAP image put inside.

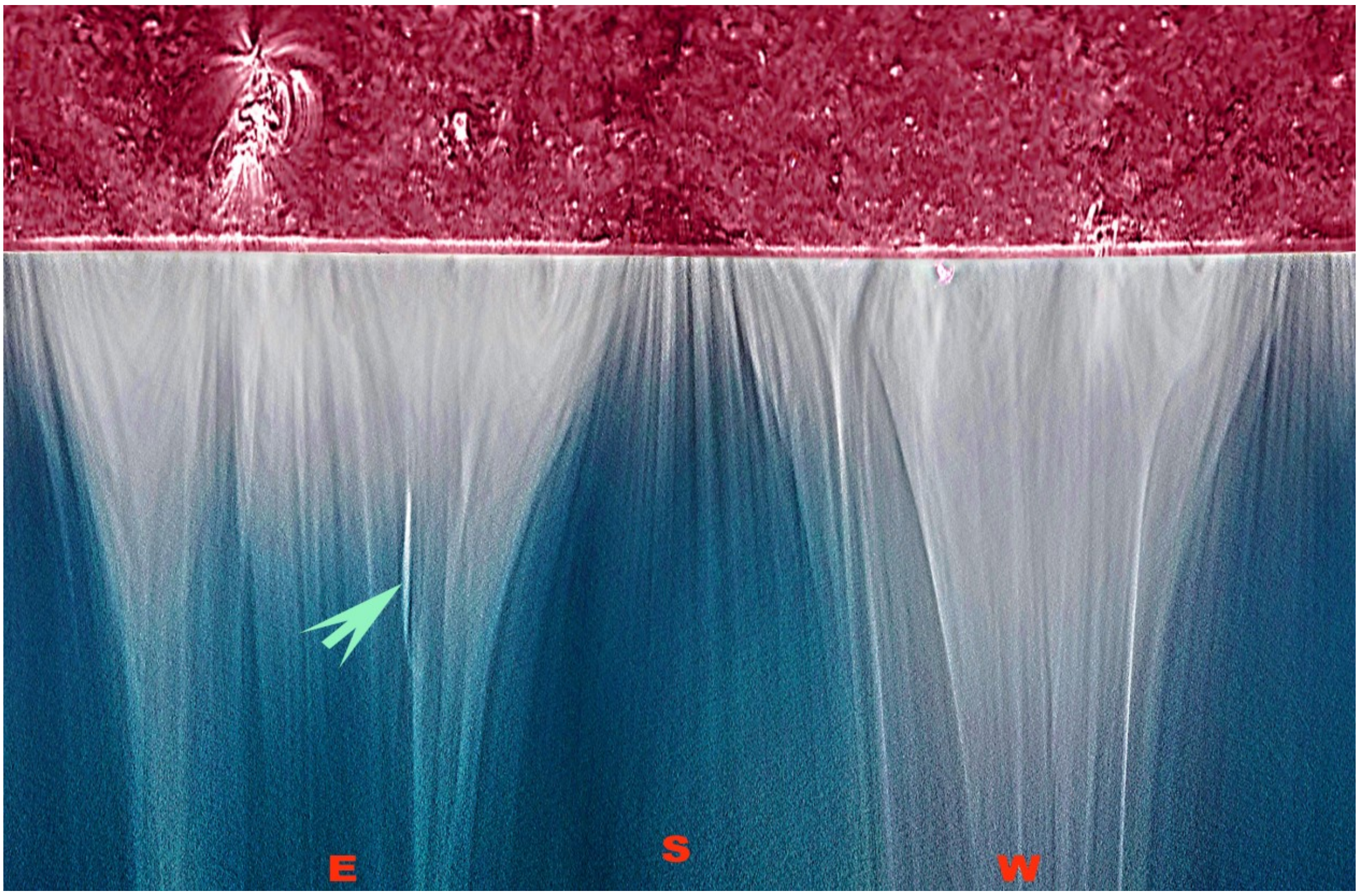

**Figure 1.** W-L composite in polar coordinate ($r$-$\theta$) made from a set of original images taken with different exposure times with the disk EUV 174 Å image from SWAP put under the limb (in red color). The top of the image corresponds to a radial distance of 0.5 $R_\odot$. The eclipse site was on the atoll Hao in French Polynesia (observer: Jean Mouette). Note the prominent feature at EES shown by the arrow that we identified as a moving blob (plasmoid) using the SWAP time sequence, see the provided movie.

During the ground-based eclipse observations, many images were collected on spacecrafts of the SoHO, the SDO (see Figure 2), the PROBA-2 (see Figure 3) and the STEREO missions.

The event started at 12:06 around the position angle 106 deg showing a width of 30 deg where a large streamer extends radially to a large distance, see the excellent discussion of he phenomenon in the paper of Habbal *et al.* (2011) and also in the paper of Pasachoff *et al.* (2011). The LASCO C2 movie illustrates quite well the flow associated with this event, the given linear speed in the CME catalogue of NRL is 206 $kms^{-1}$. It includes the propagation of a large roundish blob inside the streamer (the overlapping along the LOS of different events is always possible) and we will refer to this observation to discuss the velocities observed in the part of the corona at $r$> 2.5 $R_\odot$. Note that the STEREO mission data did not permit to directly look at the plasmoid event reported here because the LOS is not favorable as both the B and the A spacecrafts are almost in quadrature with the position of the Earth. Indeed our rather weak event occurs well above a bright inner corona structure such that the projection on the disk is difficult to detect. However, looking at the full-resolution STEREO B 195 Å channel we can see (online available movie #1) some increased emission to the left of the emerging flux region that may be the low coronal signature of the liftoff of the plasmoid region. Because this is such a weak event and it is not eruptive, it makes it difficult to say with a definite certainty and we decided not to discuss it nor to show it.

### 2.1. AIA Observation

SDO/AIA high resolution images unfortunately do not have a large enough FOV to follow the radial motion of the plasmoid but very interesting features are observed just under, using a movie specially prepared for showing this region, see movie #2.

Viewing a full-resolution movie specially prepared for us by A. Engell (private communication) of the radially filtered AIA 171 Å channel, we can see the amount of fine detail of the event resolved by the instrument. Horizontal loops are extremely dynamic throughout the liftoff of the plasmoid almost 1 day before the acceleration phase. Material can be seen in motion as early as 12:00 UT on the 10th and some material initially peels away from the solar disk and migrates into a cusp-like formation beginning around 19:30 UT on the 10th. To the south of the cusp emission from more horizontally oriented coronal loops continually unwind as outward propagation increases. Another eruptive event at 05:35 UT appears to be in front of the plasmoid liftoff. But because all this emission is off-disk it is difficult to discern what is in front and behind. Regardless, the second eruptive event does not seem to effect the propagation of the coronal loops.

We found more interesting and relevant to use specially prepared frames (see Figure 3) from processed summed images obtained by making the ratio of the AIA images obtained with the 193 Å filter (typical temperature sensitivity 2 MK) over images made with the 171 Å filter (typical temperature sensitivity 1 MK) which permits to see the off- limb loops with an increased contrast (see Figure 3). It is qualitatively showing the initial phase of the plasmoid ejection as a result of a slow dynamical evolution with a possible magnetic reconnection process.

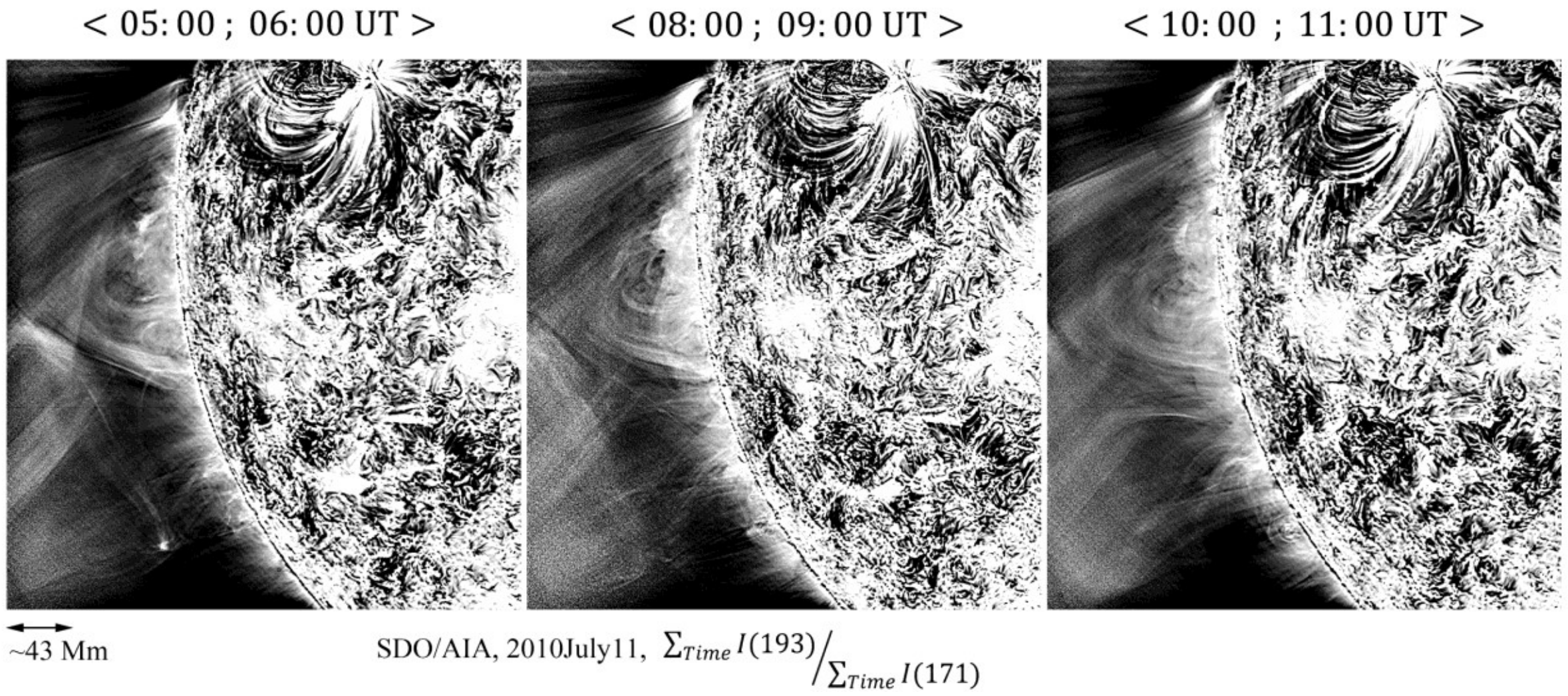

**Figure 2**. Highly contrasted summed images from selected frames of the 193/171 intensity ratio from the AIA- SDO mission. Processed images to remove the radial gradient variations and improve the contrast. Note at left the loops under the plasmoid event that we analyze, showing a cusp over revealing a possible magnetic neutral point.

More relevant for the analysis of the radial acceleration phase of the plasmoid are the results from a specially focused "eclipse" program performed with the SWAP imager of the PROBA -2 ESA mission (see Berghmans *et al.,* 2006; Defise *et al.,* 2007) for a full description of the SWAP EUV imager.

### 2.2. Swap Observation

Briefly SWAP is a solar EUV imager launched in 2009 on the PROBA-2 mission (ESA). It is using an off-axis Ritchey Chrétien telescope equipped with a new EUV detector. This type of detector has some advantages for solar EUV imaging. SWAP is providing coronal images at a 1-2 min cadence in a pass band around 174 Å. Observations allow to analyze coronal phenomena with a temperature around 1 MK. Images are of $10^3 \times 10^3$ pixel size with linear pixels corresponding to 3.17 arcsec such that the total FOV is 54×54 arcmin$^2$ which significantly surpasses the FOV of AIA imagers.

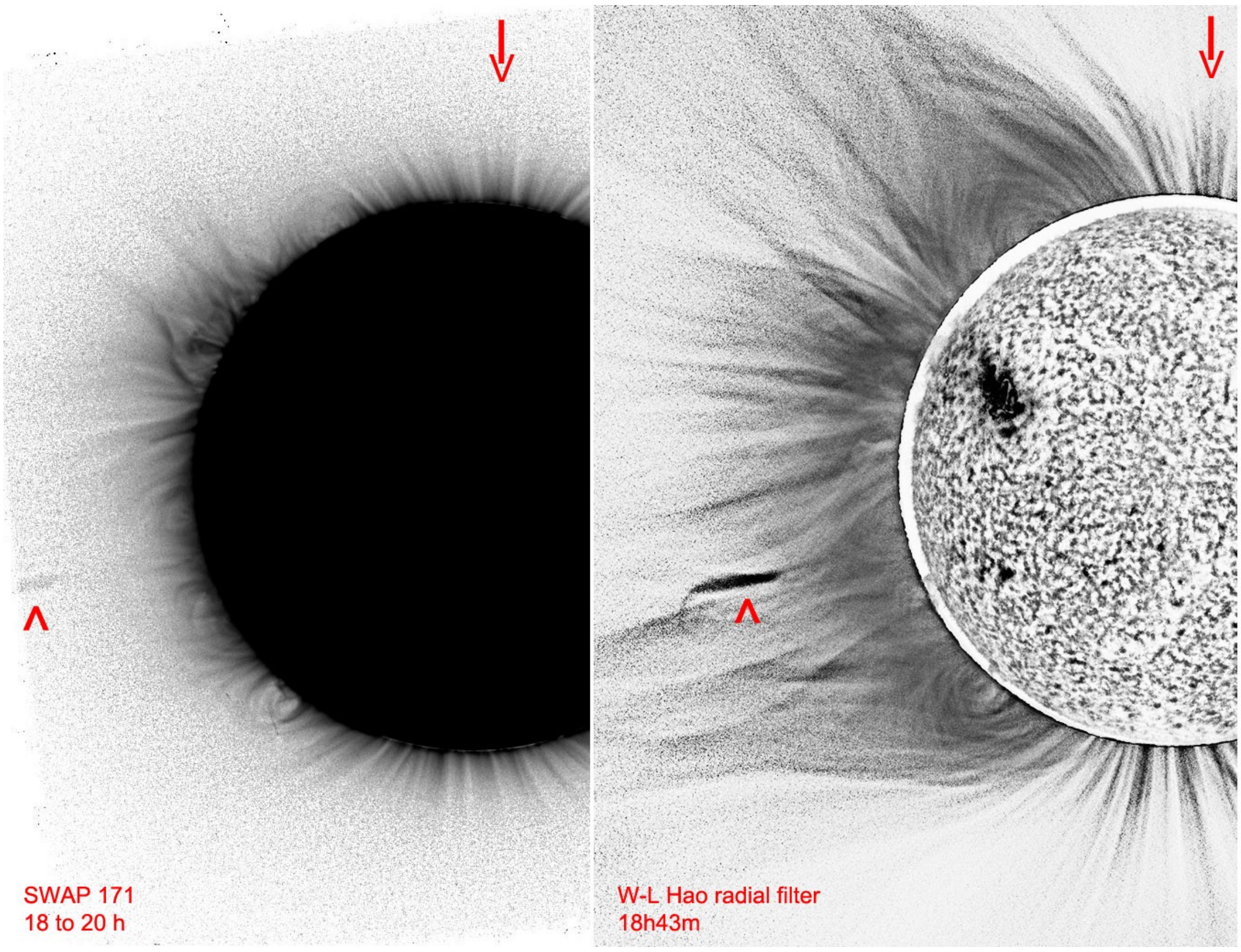

**Figure 3**. Comparison between a highly contrasted summed negative partial EUV image from SWAP (at left) and the W-L negative eclipse image (at right), same size, same field and approximately the same time with the simultaneous negative AIA 304 Å high contrast chromospheric image put instead of the Moon image. Note on the image at right the plasmoid shown with the stick and the body of the large streamer just under, at SE. This streamer is made from the superposition along the LOS of several components. At the top of each image we show with an arrow a triplet of polar plumes seen on both SWAP EUV 174 Å filtergram and W-L eclipse image near the North pole, for comparison.

For this work we used the whole day sequence of 11 July, 2010 taken with a 10 sec exposure time every minute. Images were first evaluated with the usual running difference movie which was too noisy to see the blob at the east limb (see Figure 1). A careful examination of each image, taking into account the W-L image as a reference (see Figure 3), was performed in order to remove a few "bad" images due to the rotation of the spacecraft and to the crossing of the Moon shadow during the partial phase of the eclipse. After, to improve the signal/noise ratio for the most outer parts of the FOV of SWAP images, summed images were produced from the properly aligned first generation 10 sec exposure time images of the data basis using the IRIS software (version 5.59) of Christian Buil (2011). Evaluating some tries, we decided to typically use a set of 20 elementary consecutive similar images for making a new series of optimum summed images, see Table 1. Each summed image is then displayed in a log scale, see online movie #3 and the result is processed using the unsharp masking filter of Photoshop with a 5 pixel size window, see Figure 4 for a partial frame mosaic of a sample of the resulting images. Note that this rather heavy procedure is needed for showing the analyzed feature that was not seen on SWAP images discussed in the extended papers of Pasachoff *et al.* (2011) and of Habbal *et al.* (2011) devoted to the analysis of this 2010 eclipse corona images.

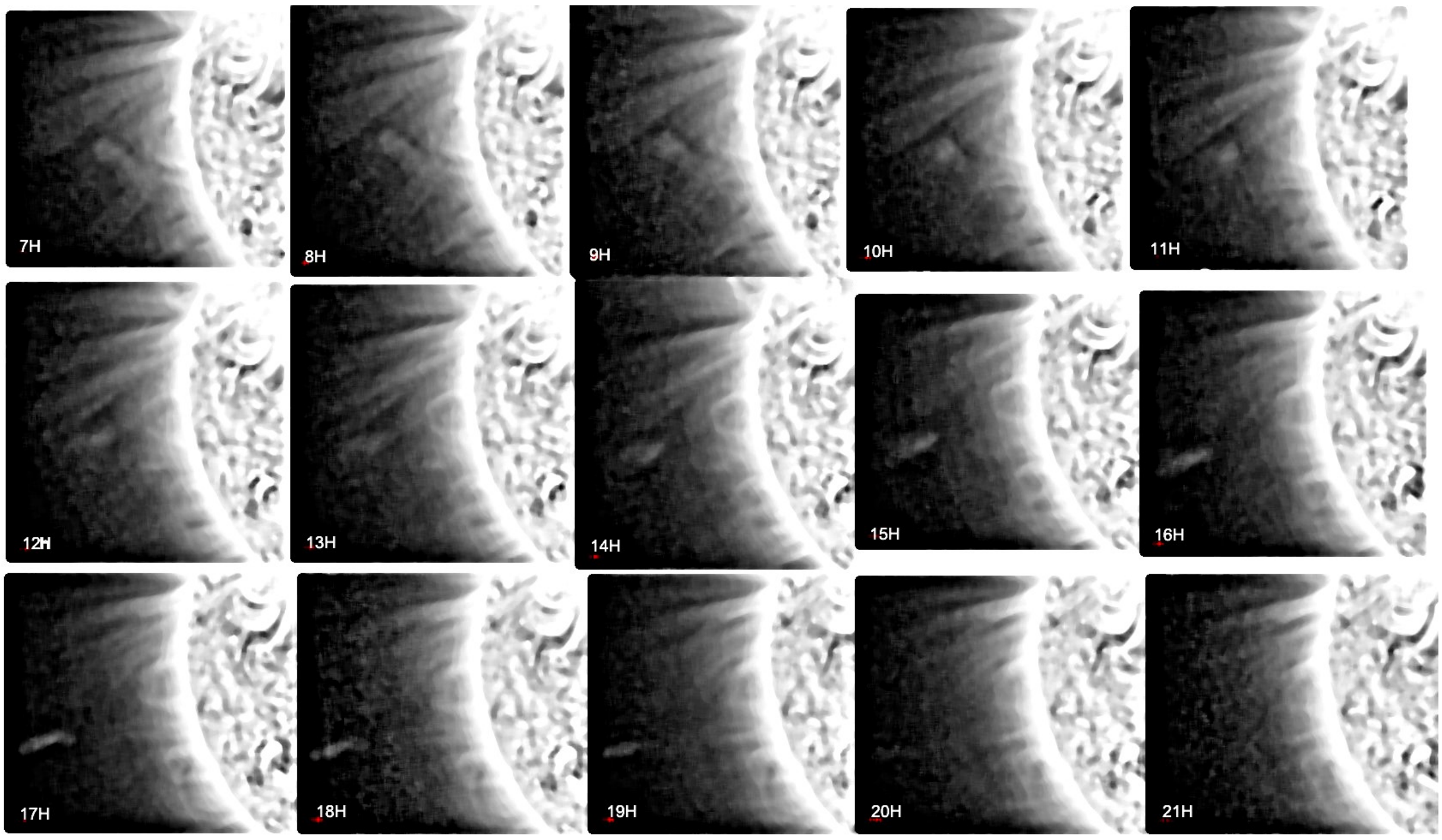

**Figure 4.** Mosaic of partial frame positive images taken from the series of processed (summed and unsharp masked) SWAP EUV images. The middle time corresponding to the set of 20 selected images used to make each composite is shown inserted at the bottom left of each frame. The contrast is increased in order to better show the blob (plasmoid). Its motion is also evaluated using the accompanying online movie #3bis.

Further the result is recorded with an increased contrast to permit the best possible evaluation of the proper motion of the plasmoid (see Figure 4).

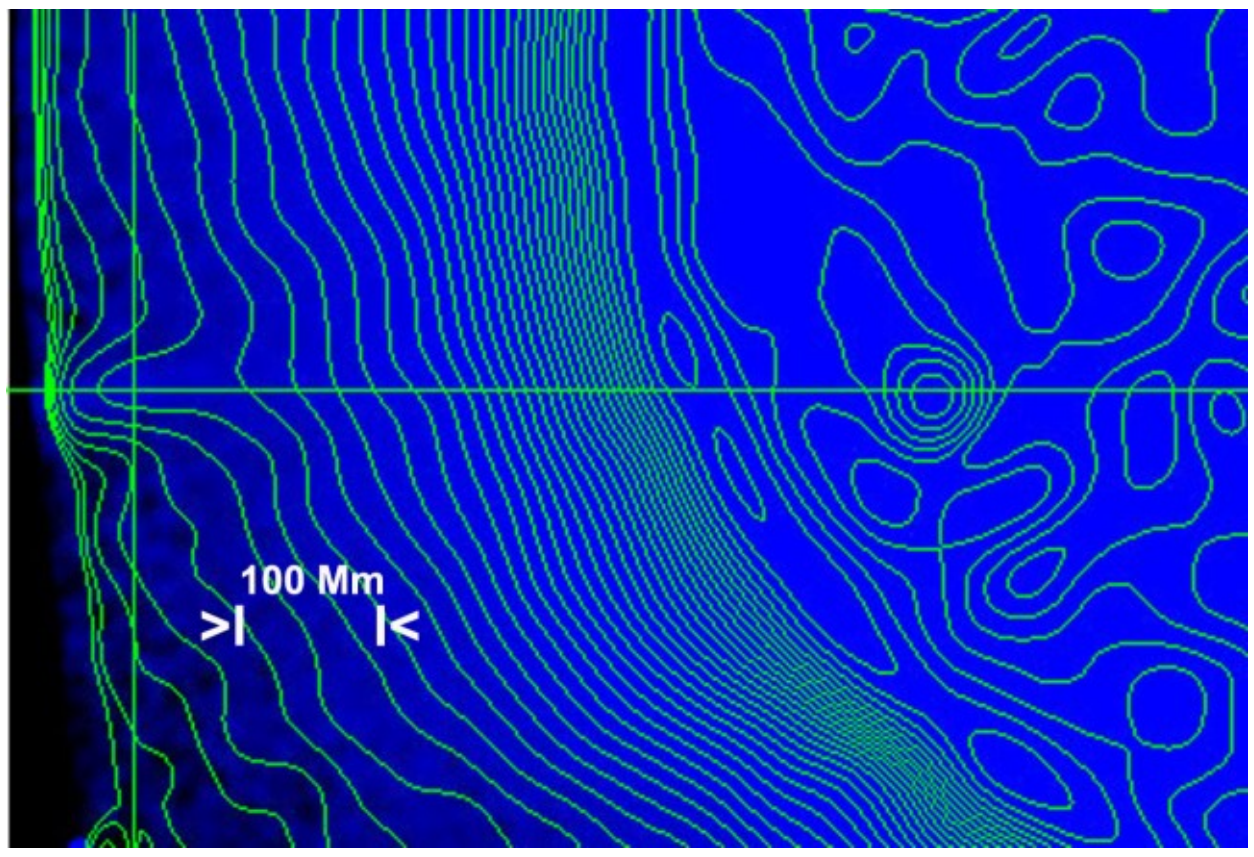

**Figure 5.** Isophote map showing the outer regions and the blob (plasmoid) from the SWAP EUV observation taken around 18h see Figure 4. Successive isophotes of the map correspond to a step of 25% in intensity levels in this averaged image, see Figure 6 for a photometric evaluation. To improve the signal/noise ratio for the blob (plasmoid), 20 images were summed to make the analysis.

It has also been used to make the greatly accelerated online movie #3bis that we present as additional material. Let us notice that the photometric part of this work is performed on the resulting summed but not on the contrasted or unsharp masked processed images of Figure 4; sometimes the results are checked by coming back to the original, not summed image.

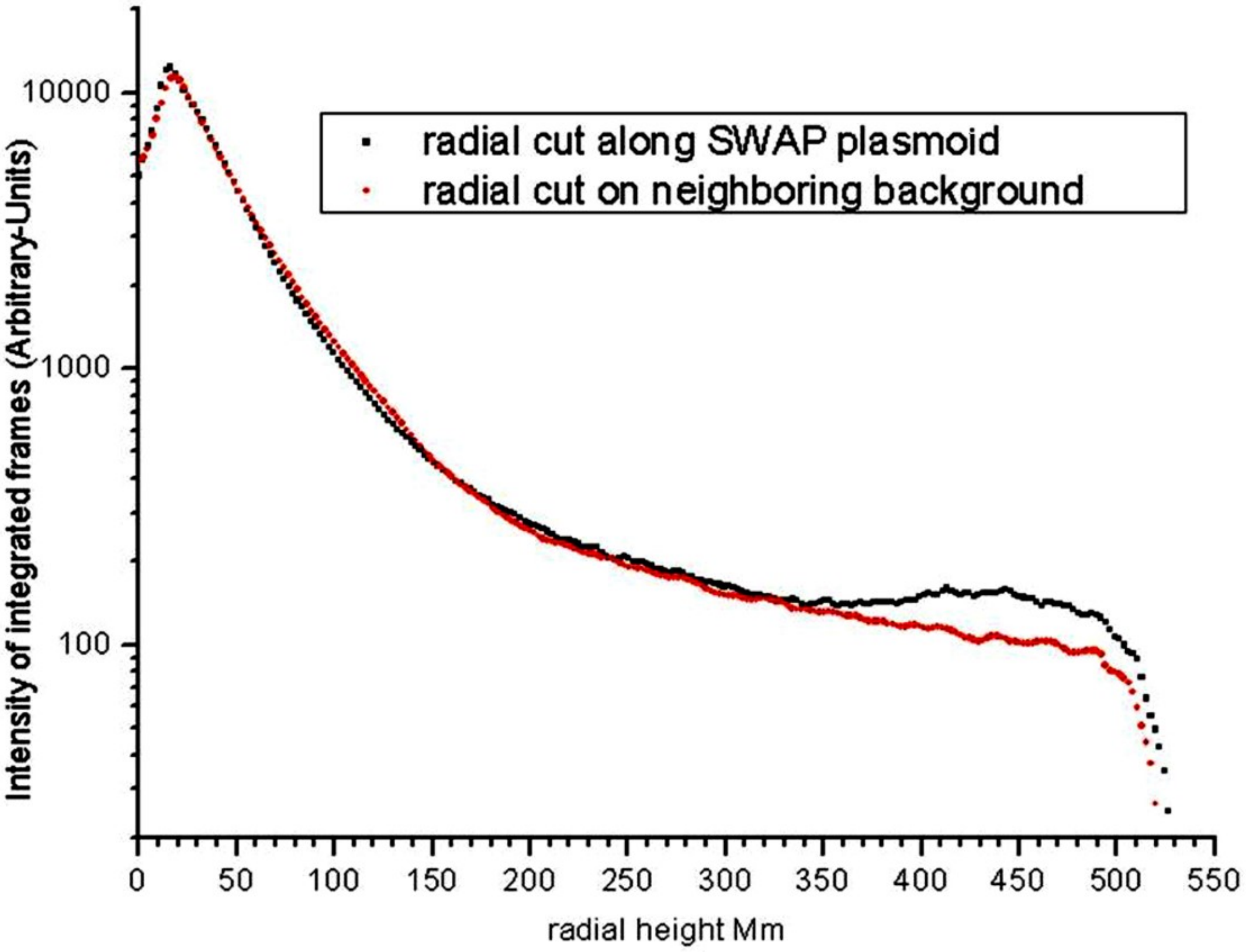

**Figure 6**. Photometric cuts along a radial direction crossing the plasmoid (in black) and along a direction very nearby in the background (in red) using the SWAP EUV summed images around 18h, see Figure 4 for a qualitative evaluation. Integration is made in the tangential direction, over ~15 Mm, in order to improve the signal/noise ratio. The stray light which is instrumental in origin, typically 30 in the same units in the outer parts, is removed.

Finally, the processed summed images were used to perform the proper motion analysis and deduce the velocity and the acceleration of the blob (plasmoid), see Figure 7. Because the plasmoid is changing of shape and shows a radial gradient along its long axis, the analysis is not precise enough, resulting in rather large error bars. However, it is clear that we deal with a definitely very slowly evolving event.

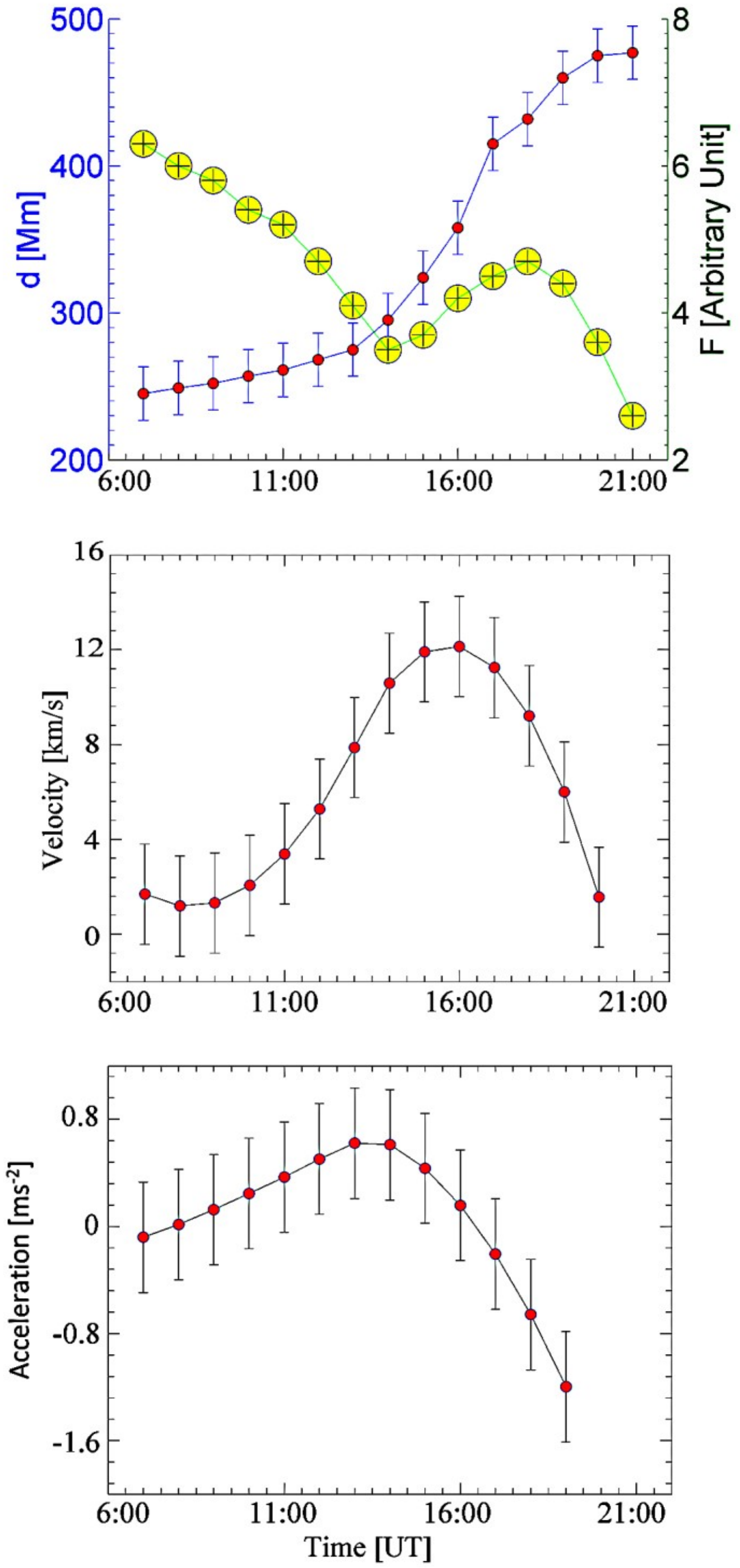

**Figure 7**. i) Proper motion (red dots) and light flux (yellow circles) variations (at top) as a function of time; ii) velocities deduced from the proper motion (at middle) as a function of space in radial direction and iii) the corresponding acceleration (at bottom) using the summed images showing the blob (plasmoid). Both the centroid and the edges of the feature are evaluated to make the measurements, giving the shown error bars.

Table 1. Datasets

| $T_{MID}$ (HOUR) | T_START | T_END | *N* | SWAP CADENCES |
|---|---|---|---|---|
| 7 | 6h24 | 8h12 | 19 | 1 IMAGE/6MIN |
| 8 | 7h00 | 9h00 | 19 | 1 IMAGE/6MIN |
| 9 | 8h00 | 10h00 | 17 | 1 IMAGE/6MIN |
| 10 | 9h00 | 11h00 | 20 | 1 IMAGE/6MIN |
| 11 | 10h00 | 12h06 | 19 | 1 IMAGE/6MIN |
| 12 | 11h00 | 13h00 | 18 | 1 IMAGE/6MIN |
| 13 | 12h00 | 14h00 | 19 | 1 IMAGE/6MIN |
| 14 | 13h00 | 15h00 | 19 | 1 IMAGE/6MIN |
| 15 | 14h00 | 16h00 | 16 | 1 IMAGE/6MIN |
| 16 | 15h00 | 17h01 | 19 | 1 IMAGE/6MIN |
| 17 | 16h00 | 18h00 | 20 | 1 IMAGE/1MIN |
| 18 | 17h00 | 19h00 | 22 | 1 IMAGE/1MIN |
| 19 | 18h00 | 20h00 | 20 | 1 IMAGE/1MIN |
| 20 | 19h06 | 21h05 | 21 | 1 IMAGE/1MIN |
| 21 | 20h00 | 21h48 | 22 | 1 IMAGE/30S |
| 22 | 21h00 | 23h03 | 20 | 1 IMAGE/30S |
| 23 | 22H00 | 23H57 | 18 | 1 IMAGE/1MIN |

**Table 1.** Data concerning the summed images series. *N* is the number of images used to get a composite in the series for making the movie. Note that only 20% of all available images were used due to a drastic selection for using the unperturbed images. $T_{mid}$ (1st column) is the adopted middle time of each set of images used to make the corresponding composite.

### 2.3. W-L Images

Eclipse images were obtained i) from a site on the HAO atoll of French Polynesia, at 18h43 UT and ii) from Easter Island, almost 1h30 min after. For the photometric measurements we used only the HAO images, although the Easter Island images are excellent and even better from the point of view of the quality of the sky, but they do not show as well the plasmoid that we are interested in (from the SWAP sequence, see Figure 4, we see the plasmoid disappearing at that time). First a composite was made see Figure 1, in order to show at best the coronal structures all around the Moon. Images rapidly taken with different exposure times (from 1/500 to 2 sec.) were assembled after properly aligning and compensating for the levels of radial intensities. On this processed composite, the elongated feature at SE looks like a detached plasma cloud and there is little doubt that the SWAP images are showing the same feature but in a more uniform environment. The impression that we deal with a moving blob (plasmoid) is confirmed when watching the SWAP accompanying movies #3 and 3bis (see the online additional material). The radial gradient of the feature, see Figure 8, is abnormal in a sense that a definite photometric "hill" is seen on the radial cut performed along the direction of the plasmoid using the original images. By subtracting the surrounding background it is then possible to compute the excess K-corona brightness of the feature and deduce the electron densities after making some assumption regarding the geometric parameters of the feature. The most critical point here is indeed the absolute calibration of our images. We used photometric profiles of very well known bright and calibrated stars of the same field as the corona simultaneously measured on our best original CMOS (complementary metal-oxide semiconductor) filtered coronal images. Here only the green channel of each image was measured. In Figure 8 the radial brightness is shown in absolute units after removing the contributions of the sky brightness and of the F-corona, following the classical methods of coronal eclipse photometry performed in W-L, see *e.g.* November and Koutchmy (1996). The details of the absolute photometry, and how the background is removed, is a rather huge topic; we already described the method in several papers dealing with the absolute photometry of preceding eclipses, see *e.g.* Koutchmy *et al.* (1978), Lebecq *et al.* (1985).

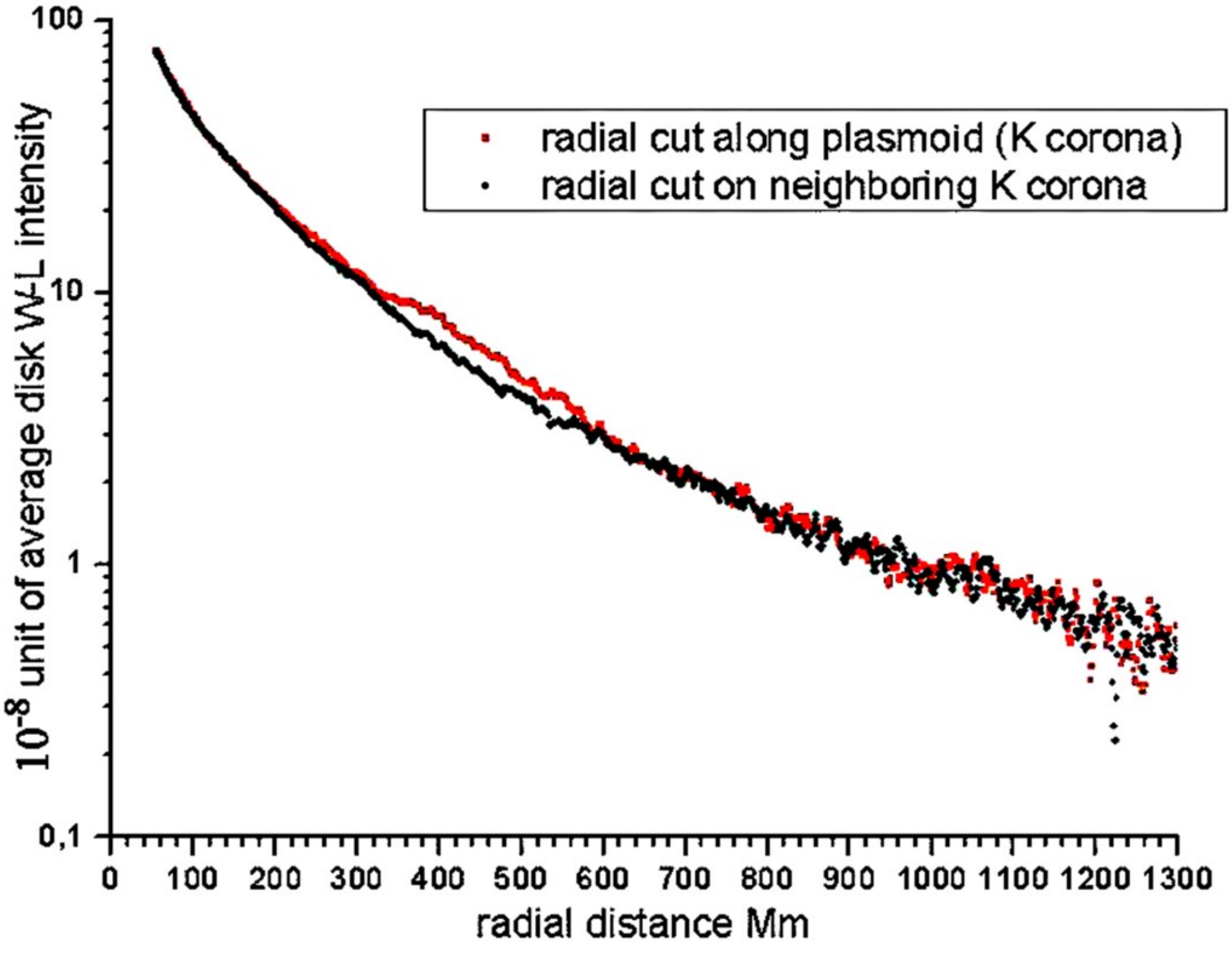

**Figure 8.** Radial photometric cuts through the blob (plasmoid) and nearby, from the absolutely calibrated image of the CMOS single image taken with an exposure time of 1/5 sec. using the green light. Absolute brightness of the K- corona shown in units of the average solar disk brightness, after subtracting the F- corona radial component. .

We further assume that the plasmoid is evolving in the plane of the sky and its column density corresponds to an integration LOS length of 25 Mm comparable to the photometric FWHM across it. Then, the deduced typical electron density is found to be $10^8$ cm$^{-3}$ at $r$=1.7 solar radius and the typical surrounding background density is there close to $10^7$±20% cm$^{-3}$, which is one order of magnitude less but it is close to the classical values in the equatorial corona.

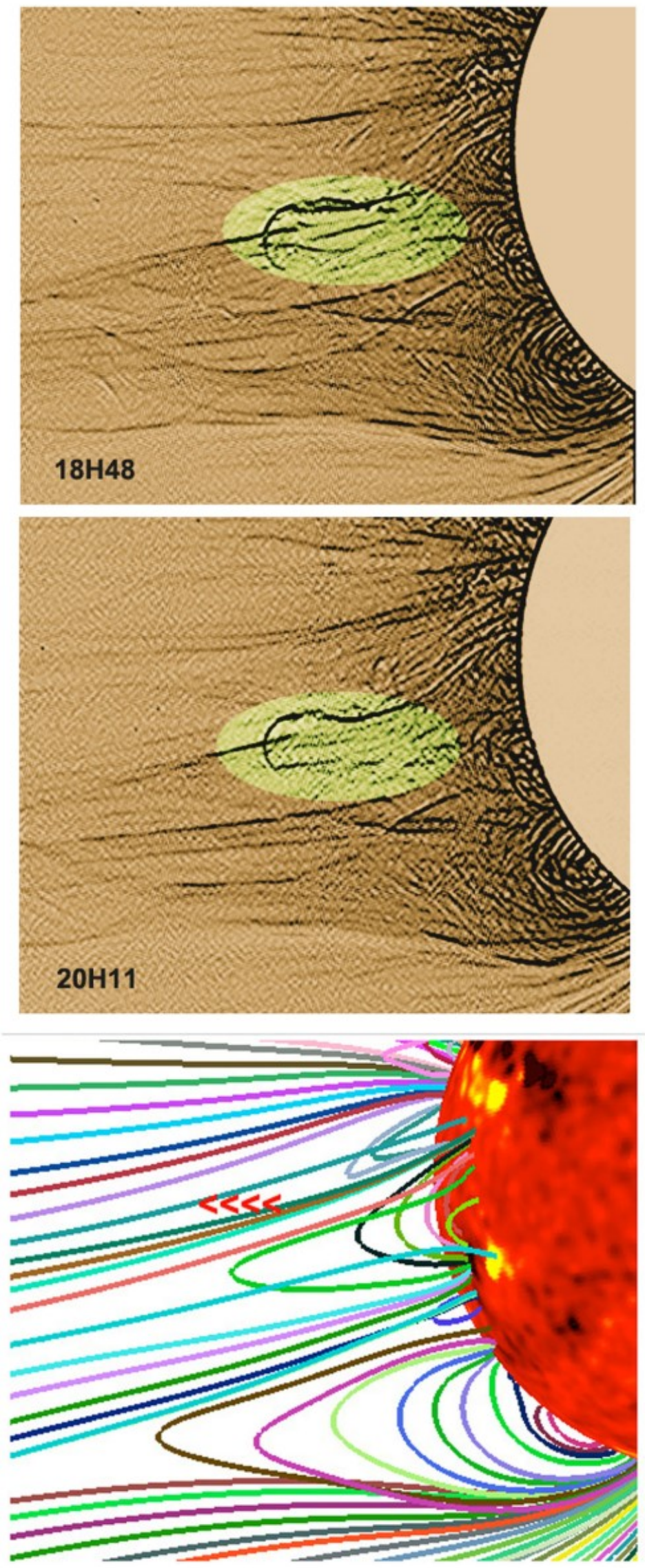

**Figure 9.** At the top, extract of the high resolution reconstructed W-L coronal images after processing (the operator Madmax was used) to show at best the coronal structure defined by small scale gradients supposed to represent the magnetic field topology; eclipse images obtained from different sites and different times are shown (observations by J. Mouette, M. Druckmuller, Ch. Nitschelm and J-M. Lecleire). At the bottom, the magnetic field lines computed from a full set of MHD equations (Z. Mikic *et al.* at Predictive Science Inc. in San Diego, USA) and from the surface magnetic field extrapolated for the date of the eclipse, to show the magnetic context around the plasmoid (arrows at left).

## 3. Discussion

### 3.1. Dynamical Properties and the Evidence of Reconnection.

After an extended image processing of just the W-L images made from two different eclipse sites to show at best the smallest coronal structure in order to "avoid" the integration effect along the LOS, see Figure 9, it has been impossible to seriously consider the question of the dynamical behavior of this blob (plasmoid) slow event.
Fortunately, from the SWAP full sequence of images (see *e.g.* the movies #3 and #3bis in the additional material) the proper motion seen from the beginning to the end of the event is easily analyzed see Figure 4.

The deduced velocities (Figure 7) are larger than the rigid rotation velocity (3 km s$^{-1}$ at the radial distances from 1.3 to 1.7 R$_\odot$) of the corona. They are much smaller than the usually adopted turbulent velocities inside the corona (of order of 20 km s$^{-1}$). These "bulk" motions are probably determined by the magnetic structure of the low beta plasma. Further out, radial velocities are growing and the plasmoid accelerates at the time of the total eclipse an after, decelerates see part iii) of Figure 7. On Figure 10 we show several type of velocities measured in this region. They correspond to different components of the streamers, in good agreement with the velocities deduced from the equation of mass continuity for a typical streamer by Koutchmy (1972). This behavior has nothing to do with the rapid acceleration of blobs observed in connection with CME events, see the recent paper of Schanche *et al.* (2016). In that sense we call it a failed plasma ejection. Note that observations of blobs and plasmoids against the solar wind flow (falling back) were already reported for the more outer regions analyzed with LASCO- C2 (SoHO) observations (*e.g.* Boulade *et al.,* 1997; Wand and Sheeley, 2002).

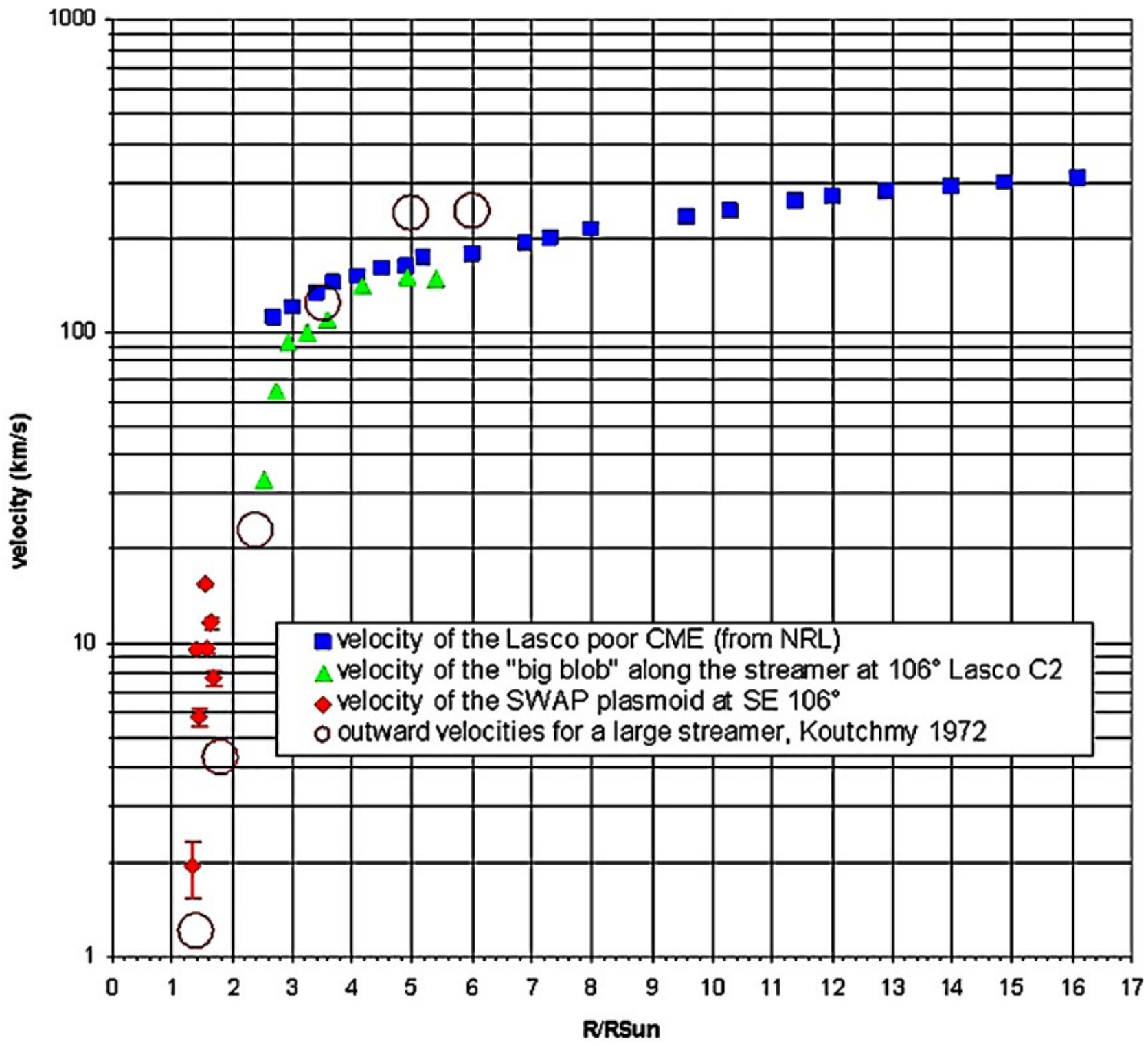

**Figure 10.** Velocities and proper motions obtained for different non simultaneous phenomena related to a perturbed streamer i) for the CME at 106 deg (LASCO observations) after 12 UT (blue squares), ii) for a well defined quasi-spherical plasma blob flying radially inside the SE streamer after the CME (green triangles), iii) for the SWAP plasmoid in the more inner corona (red diamonds with very low values). The green- blue velocities corresponding to the escaping the Sun plasma follow the typical values reported for the empirical model of a large eclipse streamer by Koutchmy (1972) (empty brown circles).

From the inspection of Figure 11, the mechanism which leaves the plasmoid to flow away could be the classical break-out mechanism proposed by *e.g.* Syrovatsky (1982) for active region loops and by Antiochos *et al.* (1989) for the CMEs. Above the loop system from where the plasmoid emerged, possibly near a magnetic null point (see Figures 2 and 11), the outer magnetic-field seems open from the inspection of extrapolated magnetic-field made using the Predictive Science Inc. MHD simulation (see Figure 9). The discussion is however biased because the simulation cannot take into account the temporal variations occurring between the limb situation and the disk observation used for the simulation. Potential field extrapolated from the surface gives small strength values at a height above $r$=1.5 R$_\odot$ see Poletto *et al.* (1975), suggesting a rather high-$\beta$ environment (Kuijpers and Fletcher, 1996). Further out, the stretched by the flow magnetic field could probably be enough to allow a scenario that has been suggested for collisionless plasma: reconnections making magnetic islands (plasmoids), as in the classical simulations of Lee and Fu (1985) and Liu and Hu (1988). Plasmoids could also be accelerated by the so called melon seed mechanism see Schlüter (1954), working for diamagnetic plasma. Other possibilities for producing a plasmoid in a low-$\beta$ environment would involve the shearing of a magnetic arcade (see Figures 2 and 11); subsequent disconnection occur when some resistivity is included in the model, like in the extended works of Mikic and Linker (1994) and Amari *et al.* (1996), made to represent a CME event.

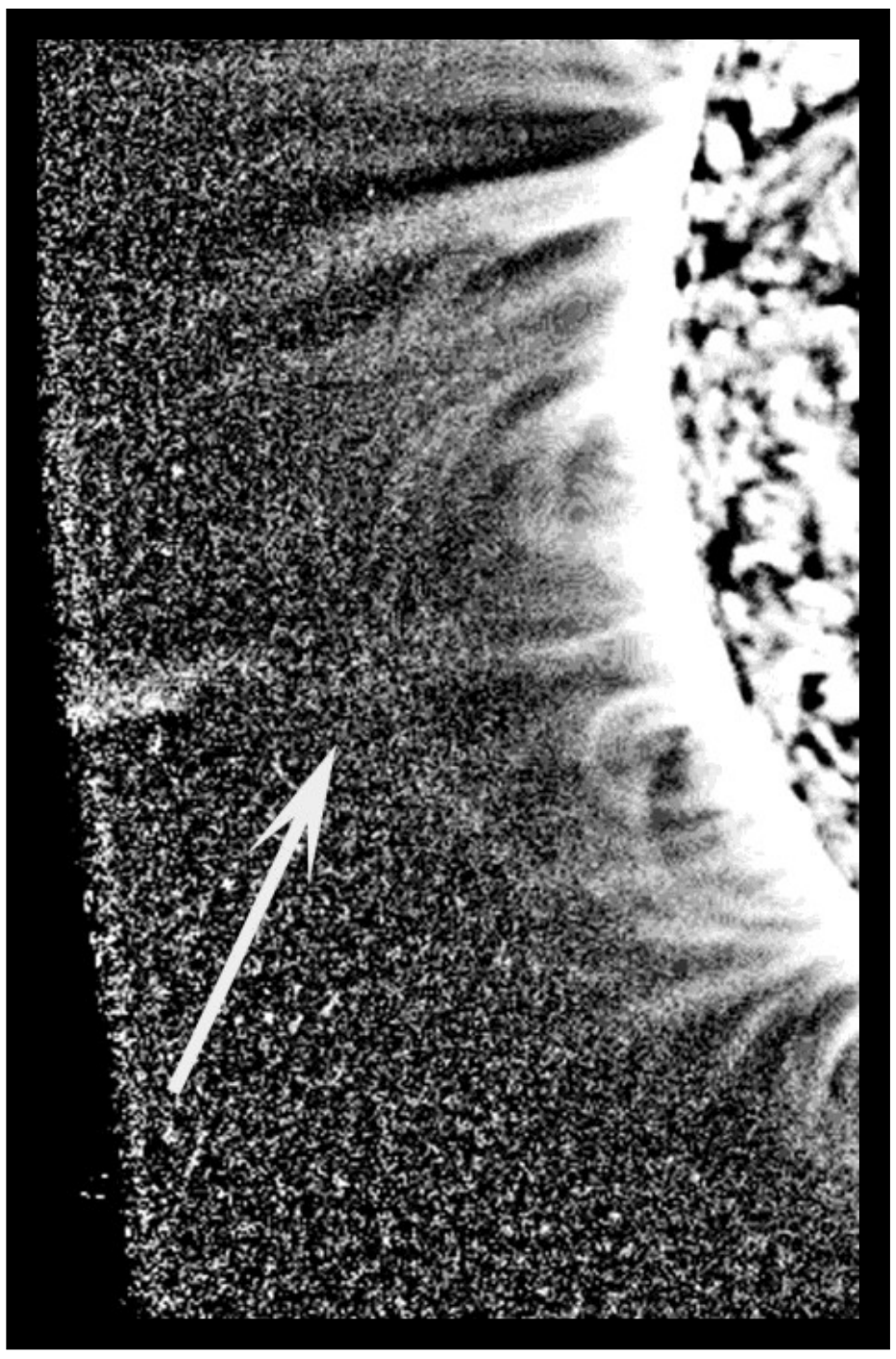

**Figure 11**. Highly processed summed over several hours image from the SWAP EUV low temperature (1 MK) plasma observations to show a possible X- type neutral point under the blob (plasmoid). This is in striking difference compared to the high resolution W-L eclipse images that reflect the plasma densities at different temperatures (see Figures 1, 3 and 9).

### 3.2. The Mass Loading Question

The supply of material to the slow wind is clearly related to the radially extended coronal streamers (Figure 1). A "weak" and narrow CME was incidentally recorded several hours before the appearance of the blob (plasmoid) in the nearby layers (see the discussion given in Habbal *et al.,* 2011). Stealth CMEs (see Fa *et al.,* 2010) are also recorded with the W-L coronagraphs. From the statistical analysis of CMEs and dynamical events seen with the LASCO C2 coronagraphs, for more than 1 solar cycle, collected by Robbrecht *et al.* (2009-b) with their Cactus tool, it is clear that the number of events under $r$=2.5R$_\odot$ should drastically be increased. We can assume that at lower heights where the plasma density is much higher and field lines are tighter, more events would be observed. Among them failed ejections with plasmoids similar to the case described here are good candidates to provide mass. W-L systematic good resolution observations in this region, say from $r$=1.3 R$_\odot$ to $r$=2.5 R$_\odot$, are cruelly needed. Our analysis shows that EUV observations using the 174 Å channel, like SWAP observations are, could be a good substitute to W-L eclipse observations such that they may help to resolve the problem of the slow wind provided they would have a larger FOV to better show the outer parts of the inner corona near 1 R$_\odot$ from the surface, without the need to perform a summing of many frames. Obviously routine W-L observation of the inner corona near 1 R$_\odot$ from the limb will provide an evan better solution to the mass loading question as proposed by the PROBA-3 mission (Lamy *et al.,* 2008). Accordingly, failed ejections under 1 solar radius from the surface will not be neglected in the future as a part of the dynamical phenomena that eventually, participate in the mass loading of the corona, including the feet of streamers responsible for the slow wind.

### 3.3. The Resonance Scattering Component of the Line Emission

It is also necessary to better understand what is the origin of the recorded radiations in the 174 Å filtergrams at $r$>1.5 R$_\odot$, *i.e.* in the outer corona. Since a long time it has been suggested that the resonance scattering (RS) could be dominating the collisional excitation (CE) for lines with high oscillator strengths, like those in the 174 Å pass band of SWAP, see Fawcett and Mason (1991). The RS radiation was repeatedly advocated in the recent papers of Habbal for explaining her Fe XI deep near IR eclipse images at 7892 (see Habbal *et al.,* 2011 and references inside). In his early seminal paper Allen (1975) computed the respective contributions of RS and of CE and came to the conclusion that RS would dominate in the more outer layers unless some special effect due to the so-called heterogeneity factor is introduced. We also note that Doppler velocities of the plasmoid would barely surpass the values of velocities corresponding to observed proper motions and, accordingly, in our case the Doppler dimming effect is almost negligible for the RS effect. Indeed Schrijver and McMullen (2000) already considered the problem for the case of TRACE observations collected with their 171 Å channel, close to the one of SWAP. They showed that the RS is responsible for the background haze seen in the high-resolution TRACE images of the quiet corona. We have a quite similar but fainter aureola effect in SWAP filtergrams, see *e.g.* Figure 5. The measured level of the background is not only of coronal origin. Auchere and Artzner (2004) discussed the problem of the stray light polluting the EUV filtergrams and although some progress was made in the coating of the SWAP mirror, the scattered light level has to be measured and only some preliminary evaluation has been made from the partial eclipse images of 11 July, 2010. It is important to improve this evaluation before claiming what is the origin of the EUV radiation measured on the plasmoid with, hopefully, a determination of its temperature see the recent discussion in the Goryiev *et al.* (2014) paper. This is partly done with the photometric work devoted to the comparison of the intensities of the blob (plasmoid) in both W-L and EUV.

**Acknowledgements** This work has benefited from the SWAP G-I program developed at the ROB; A. De Groof, D. Berghmans, B. Nicula and D. Seaton helped with the preparation of an earlier preliminary report of the SWAP observations. We also thank J. Mouette, J-M. Lecleire, Ch. Nitschelm and especially M. Druckmuller for providing excellent eclipse images, and L. Golub, V. Slemzin, A. Engel, B. Filippov and Z. Mikic for useful discussions and meaningful suggestions.
Eclipse observations were supported by CNES (France); we thank J-Y. Prado and P. Martinez for their interest and implication in our work. More discussions and suggestions during the course of this work by Andrei Zhukov, Laurent Dolla, Frederic Auchere, Jean-Claude Vial, Guillaume Aulanier, Jacques Dubeau, A. Urnov, E. Podladchikova, Y-M. Wang and others were greatly appreciated. E.T. are also grateful to the Iran National Foundation INSF. The SDO and the STEREO missions are space missions by NASA.

**Disclosure of Potential Conflicts of Interest** The authors declare that they have no conflicts of interest.

**Additional (accompanying) material:** online movie

- # 1- The STEREO B 171 Å disk and radially filtered movie of 11 July, 2011,
- # 2- AIA radially filtered movie of July 11, 2011 from 13UT to 21 UT,
- # 3- The full FOV movie made from a series of summed over 2h SWAP EUV images with a cadence of 1h from 7h to 23h shown in log scale,
- # 3bis- As movie #3 but using an enhanced contrast to display each frame.